# Closing the Loop: Integrating Material Needs of Energy Technologies into Energy System Models

Célia Burghardt
INATECH, University of Freiburg
Freiburg i. Br., Germany
celia.burghardt@inatech.uni-freiburg.de

Mirko Schäfer
INATECH, University of Freiburg
Freiburg i. Br., Germany
mirko.schäfer@inatech.uni-freiburg.de

Anke Weidlich
INATECH, University of Freiburg
Freiburg i. Br., Germany
anke.weidlich@inatech.uni-freiburg.de

*Abstract—* The transition to a climate-neutral energy system demands large-scale renewable generation expansion, which requires substantial amounts of bulk materials like steel, cement, and polymers. The production of these materials represents an additional energy demand for the system, creating an energy-material feedback loop. Current energy system models lack a complete representation of this feedback loop. Material requirements of energy system transformation have been studied in a retrospective approach, not allowing them as a consideration in system design. To address this gap, we integrate bulk material demand and production as endogenous factors into energy system optimization using PyPSA-Eur. Our approach links infrastructure expansion with industrial energy needs to achieve a minimum-cost equilibrium. Applying this model to Germany's transition to climate neutrality by 2045, we find that accounting for material needs increases annual bulk material demands by 3–9 %, shifts preferences from solar to wind and from local production of hydrogen to ship imports, and shows distinct industrial process route choices. These findings suggests that energy-material feedbacks should be considered in energy system design when moving to more domestic production of energy technologies.

*Index Terms*--Energy-material nexus, energy system modelling, industry defossilization.

## I. Introduction

Meeting climate neutrality targets in the energy sector requires expanding large capacities of renewable technologies. This drives substantial demand for both critical raw materials and bulk materials such as steel, cement, and polymers [1]. For energy-intensive bulk materials, this creates a bidirectional energy-material feedback loop: energy is needed to produce materials, and materials are needed to build energy technologies [2]. Unless met entirely through imports, these material demands can influence optimal system design, particularly in resource- and emissions-constrained scenarios where small changes in demand may cause significant cost increases. Currently, many technologies essential for a net zero European energy system, such as solar panels and fuel cells, are largely imported [3]. However, the EU's Net-Zero Industry Act aims to localize value chains to reduce dependency and enhance resilience [3]. These goals underscore the importance of analyzing energy system design with full energy-material feedbacks. Most research focuses on critical materials ([4], [5], review by [2]) which are relevant due to their supply risks and limited availability [6]. This is modelled by supply constraints or multi-criteria optimization (e.g. [5]). In contrast, the challenge with bulk materials is that they are required in large volumes—e.g., wind turbines are composed of 70–72 % concrete and 24–25 % steel [1], and their production is highly energy- and emissions-intensive [7], creating additional challenges for climate-neutral transitions. Considering the effect of bulk material demands in energy system models thus requires modelling energy demand and emissions related to their production. Material needs of the energy transition have been quantified with retrospective approaches, multiplying technology expansion from fixed energy scenarios with material intensity factors to estimate demand (e.g., [4]; review by [2]). This approach does not allow consideration of material needs in the energy system design. Some integrate industrial production to reflect industry energy use but treat material demands as exogenous (e.g., [8], [9], [10]). Only a few studies capture full feedbacks by endogenizing material demand and supply, but they focus on a single material such as lithium [11] or copper [12], or include bulk materials but with limited energy system resolution [13]. We propose a novel approach to integrate bulk material demand and production into energy system optimization, explicitly modeling feedbacks in both directions (see fig. 1). Our approach extends existing reasearch by introducing material demand as an additional criterion in technology selection (e.g., wind turbines, direct air capture or heat pumps), and by co-optimizing industry process selection to provide these materials domestically. The main objective is to find out whether domestic material production for the energy system transformation leads to a different optimal system design than imported technology and materials. To achieve this, we enhance the open-source energy system model

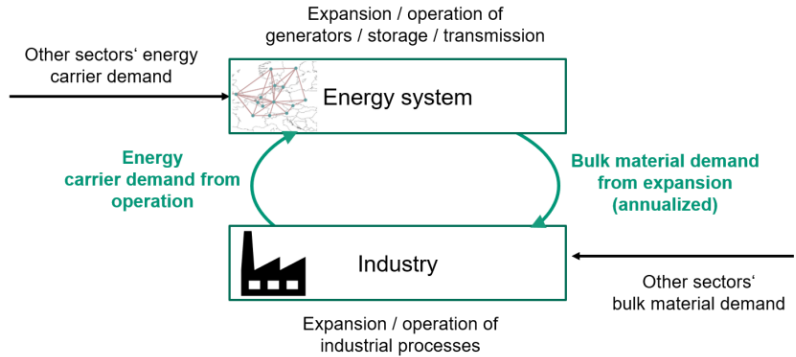

Fig. 1. Feedbacks between industry and energy system in the optimization. Material-to-energy feedback: energy carrier demand from operation of industry processes (all scenarios). Energy-to-material feedback: material demand from energy technology expansion (only domestic production scenarios). Other demands include demands from other sectors, e.g. buildings and transport. The increased material demand is only included in the increased-material reference (100 %, not linked) and set to the material demand of the scenario with full feedback (100 % domestic). For separate representation of cross-sectoral links in scenarios see appendix fig. 7.

PyPSA-Eur [14] with two modifications. First, we directly link material demands to energy infrastructure expansion. Second, we model industrial processes for bulk materials, making production routes and fuel choices endogenous. We apply this feedback-loop model for Germany's energy and industrial transformation to climate neutrality by 2045. In a reference scenario, material demands linked to energy system expansion are not considered, corresponding to the assumption that 0 % of materials are produced domestically in Germany. The other scenarios stepwise increase the domestic production share to 25, 50, 75 and 100 % respectively. In an additional reference scenario, the material demands are increased to the level of the 100 % domestic scenario, but without linking material demands to technology expansion (100 %, not linked). The feedbacks included in the different scenarios are shown in fig. 1. The following section (II) describes the optimization problems of the compared scenarios. Section III compares the optimization results of the different scenarios regarding material demands and technology choices in industry and energy system. Main findings, limitations and future work for our study are discussed in the section IV.

## II. METHODOLOGY

We develop an optimization model of the energy-intensive industry, comprising options for producing the bulk materials steel, cement, and high-value chemicals HVC, as well as for providing industrial process heat at three temperature levels. This model is coupled to the energy system model PyPSA-Eur [14] for Germany and countries with a direct power transfer link. A more detailed description of the industry and energy system models as well as a full mathematical model formulation is given the appendix in section IV.

The objective function minimizes annual system costs for expansion and operation of technologies, under emission and resource caps, in an overnight (no pathway), greenfield (no current technology capacities considered) approach for the year 2045. System costs consist of total costs of the industry sector and the energy system. The decision variables are the expansion and operation of industry processes, generators, transmission, and conversion technologies, and storage, summed for all nodes of the network and time steps of the one-year horizon. Costs consist of annualized investment costs, operational costs, and costs for consuming resources. The operational costs of the industry processes include raw material costs and exclude energy carrier costs, but industrial energy carrier demands induce costs in the energy system to provide the energy carriers. Equally, in the domestic production scenarios, the capital costs of technology expansion do not include costs for steel, cement and HVC, but demand for these materials induce costs in the industry sector to produce them (and costs in energy system for providing related industrial energy demands). This avoids double-counting of material costs addressed in [2].

In the scenarios with partial domestic production (25, 50, 75 % domestic), capital costs only exclude the share of domestically produced materials. In the reference scenarios, no material demands of energy system technologies are domestically produced, thus capital costs include bulk material costs.

In all scenarios, industry and energy system operation are connected through the energy balance as schematically shown in fig. 1. It defines for each energy carrier, that generation, storage, conversion, and transmission within the energy system must meet the exogenous demand plus the endogenous consumption by industry and conversion processes. The exogenous demand consists of demand from the non-industry sectors (e.g., buildings, transport) and aggregated demands of further, not explicitly modelled industry subsectors (e.g., pulp and paper, food processing, etc., based on 2021 data from the JRC-IDEES database [7] and from [15]). Electricity and hydrogen can be exchanged between the countries when networks are expanded in the optimization, and hydrogen ship imports are enabled at a price of 84 C/MWh.

The material balances ensure that demand for each material (steel, cement and HVC) is met by industrial production. In the reference scenarios, an exogenous material demand is set. In the other scenarios annualized material demands per capacity build-up of energy system technologies must additionally be served. Material demands are annualized by distributing them equally over the lifetimes of the technologies.

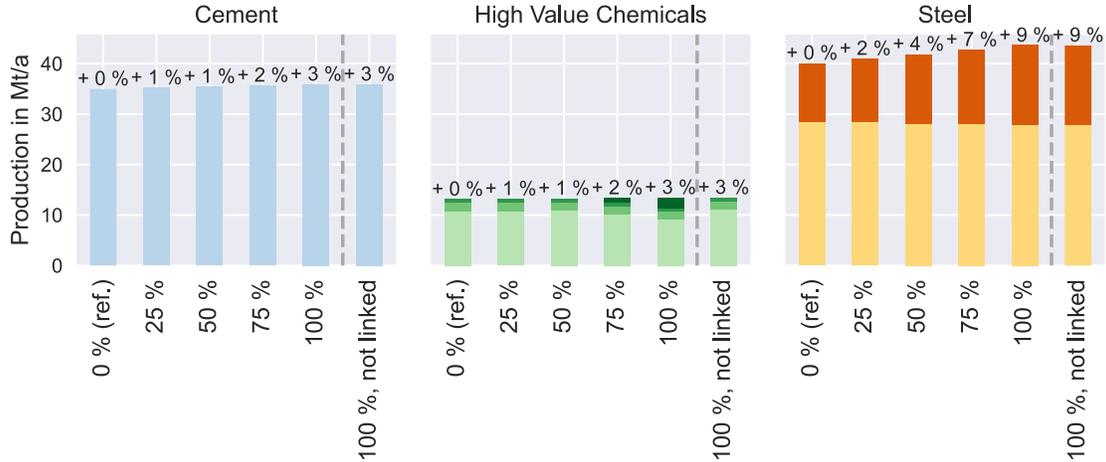

Fig. 2. Material demand and production by process route at the German node in 2045, by scenario. The scenarios vary in the percentage of domestically produced materials (here: domestically = in Germany). The numbers on the bars describe the increase in material demand compared to the reference scenario. The dashed line separates the increased-material reference (100 %, not linked) which differs in the optimization problem: material demands are set exogenously, matching the levels of the 100 % domestic scenario, but are not induced by the expansion of technologies.

This corresponds to the assumption of constant rebuilding of technologies after 2045. The per-capacity material demands of the technologies are shown in the appendix.

Limits for resource consumption and emissions are set jointly for industry and energy system, thus they compete for certain resources, such as biomass, waste, and $CO_2$ storage capacity, and waste. The emission limit is set to zero, corresponding to the limit of the German climate law for 2045.

### III. RESULTS AND DISCUSSION

#### A. Material demands and production

Fig. 2 shows the resulting total demand for cement, HVC and steel in the different scenarios and their resulting respective process routes. Material demand increases endogenously as energy system technologies expand according to domestic production shares. The increased-material scenario (100 %, not linked) sets material demands exogenously, matching the 100 % domestic scenario, but without linking demand to technology expansion. Compared to the reference (0 % domestic), 100 % domestic production increases material demands by 3 % (cement), 3 % (HVC), and 9 % (steel). Steel demand is most affected due to its high share in energy system technology materials (see material factors in appendix). Production routes remain unchanged for cement (low-clinker mixtures with carbon capture) and steel (recycling up to limits set by steel scrap availability, and H2 direct reduction). HVC production relies on chemical and mechanical recycling up to the secondary material limit. Methanol-to-olefins dominates in the 0–50 % local production scenarios, while steam cracking with synthetic naphtha (from Fischer-Tropsch) partially replaces it in the 75–100 % scenarios. The increased-material reference scenario (100 %, not linked) continues to rely entirely on methanol-to-olefins for primary production, like in the 0 - 50 % domestic scenarios. Resulting process heat technology choices are the same for low- and high-temperature heat in all scenarios, but differ in the medium-temperature range. For low-temperature heat, industrial heat pumps are used in all scenarios. For medium temperature heat, around half comes from furnaces fueled with bioenergy plus carbon capture and storage (BECCS) and half from H2-fueled furnaces in the reference scenario. In the 100 % domestic scenario, the ratio shifts towards biomass furnaces with BECCS (78 %). In the 100 %, not linked reference, this ratio shift towards biomass furnaces is less pronounced (60 %). High-temperature heat is entirely provided by biomass furnaces with BECCS in all scenarios, with total heat demand increasing as domestic material production rises.

#### B. Energy system

The energy system must meet additional demands induced by increased local material production. This results in a cost-minimal equilibrium, as technology expansion further induces material demand. Fig. 3 shows expansion for electricity generation across scenarios. Instead of additional generator expansion, most of the additional energy demand is met through H2 imports by ship (fig. 4). Renewable capacities remain stable for most technologies or even decreases for utility solar, which has a high material demand relative to capacity factors (see material factors in appendix). For dispatchable generators, H2 turbine and fuel cell capacities increase, while gas and biomass combined heat and power plant capacities (CHPs) decline with increasing shares of domestic production. This also reduces heat supplied to district heating networks from CHPs.

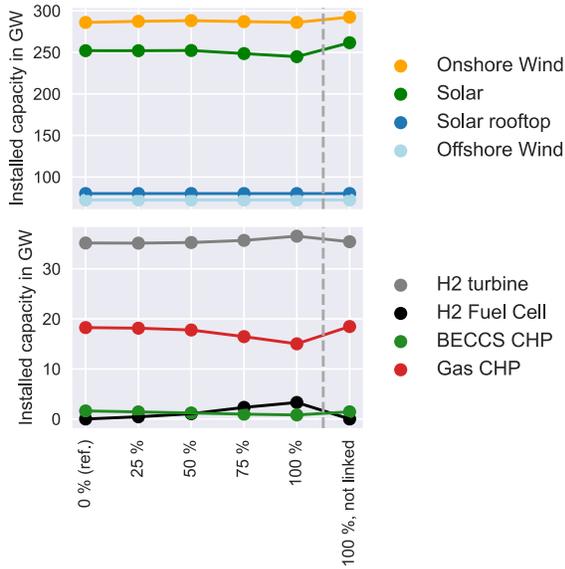

Fig. 3. Installed capacities by scenario for renewable generators (top) and dispatchable generators (bottom). Note: different y-axis scales.

Instead, heat is increasingly sourced from waste heat from Fischer-Tropsch and fuel cells in the 75 % and 100 % domestic scenarios, along with
expanded gas boilers, heat pumps, and direct electric heaters. Compared to the increased-material reference scenario, the 100 % domestic scenario results in lower overall generator capacity expansion and greater reliance on H2 ship imports (36 TWh compared to 27 TWh). H2 ship imports come from outside the modelled system and do not induce a material demand in the modelled system; instead, the material costs are reflected in the import price. This leads to reduced deployment of material-intensive solar and slightly lower onshore wind capacity. Additionally, less CHP capacity is built, with increased deployment of H2 turbines and fuel cells.

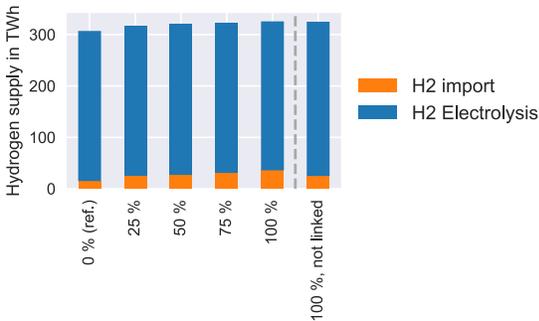

Fig. 4. Hydrogen supply by scenario.

*C. Energy-to-material feedback*

The annualized per-technology material demand in the 100 % domestic scenario is compared to the material demands of the system design in the 100 %, not linked scenario in fig. 5. In the latter scenario, material demands for technologies are not a consideration in energy system design. Thus, this comparison allows to assess the preference shift in technology choice when their materials demands are linked, comparing additional endogenous material demands linked to technology expansion versus additional exogenous material demand at the same level but not linked to technology expansion. The 100 % domestic scenario builds a more material-efficient system. For all three materials, total demand is higher in the exogenous material scenario, particularly for steel (additional 94 kt / year). This is primarily driven by increased deployment of utility solar. Additionally, more onshore wind expansion, BECCS and gas CHPs are built. These technologies drive higher material demands than the heat pumps, H2 turbines and fuel cells preferred in the 100 % domestic scenario. The amount of materials contained in technologies at the German node in the 100 % domestic scenario is shown in fig. 6, in comparison to the amount of material demand calculated for today's installed solar and wind generators in Germany. The 2045 amounts, compared to 2015, increase by factor 6 (steel), 5 (cement) and 9 (HVC), mostly used in onshore wind, solar and offshore wind generators.

IV. CONCLUSION

We present a method to integrate energy-material feedbacks into energy system models. To achieve this, we extended the sector-coupled PyPSA-Eur model by incorporating (1) an industry module representing production routes for steel, cement, and HVC, along with process heat technologies, and (2) material factors that endogenously link technology expansion to material demand. This approach addresses a gap identified in previous research [2]. We apply this method to a model of Germany and countries with a direct power link. Our results reveal two key effects of incorporating a full energy-material feedback loop in optimization. First, technology expansion drives bulk material demands by 3–9 %, which drives higher energy demand. Second, compared to a scenario with exogenously increased material demands, endogenous material feedback shifts technology preferences from solar to onshore wind and from electrolysis to H2 imports. Further technology choice differences occur in HVC production, process heat generation, and dispatchable electricity supply, leading to a system design with reduced material needs. Our findings emphasize the importance of accounting for energy-material interactions in cost-optimal system design. The method presented in this paper has several limitations. First, material demands for technology construction are distributed evenly over a plant's lifetime, an assumption that holds only if technologies are replaced at a constant rate. Second, the adjustment of technology capacity costs to exclude bulk material expenses is based on their current market prices, but the actual proportion of capital costs attributed to bulk materials may be different and change over time. Lastly, due to data limitations, material factors for some technologies were unavailable and were approximated using values from comparable technologies, and we did not include sub-technologies (e.g. wind turbines with or without permanent magnet as in [5]), assuming that the effect on structural materials is limited. Our study can be extended in several ways. First, we apply a greenfield and overnight (no pathway, no current technologies) optimization for a climate-neutral system. While this represents energy-material feedbacks in the optimal system design, it cannot represent the material requirements before 2045 to build up the climate-neutral system. Future work could apply our method in a pathway study. Second, our analysis focuses on bulk materials as structural components of energy technologies. While steel, cement, and HVC are the most energy-intensive materials in Germany, further research could expand to other materials such as copper and aluminum, and additional technologies like grids and electric vehicles. Third, our study focuses on Germany as a proof-of-concept. While effects are already visible at this national scope, extending the model to European level could offer valuable insights, allowing to consider different local conditions for renewable generation and exchange of energy and bulk materials between countries.

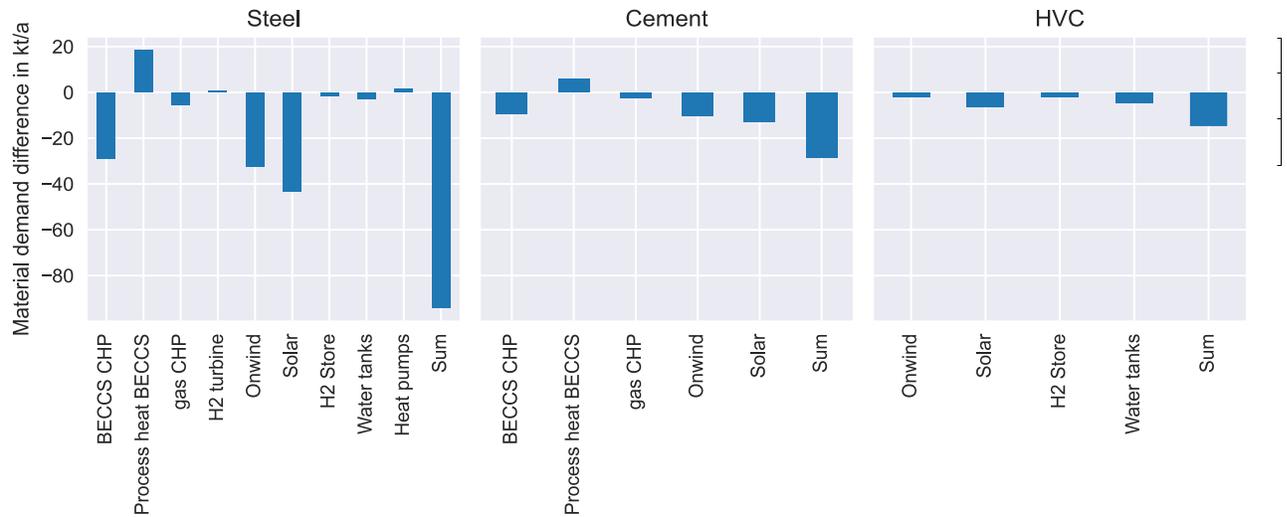

Fig. 5. System material demand by technology of the 100 % domestic scenario minus (hypothetical) system material demand of the 100 %, not linked reference. Negative values mean higher material demand in the latter scenario. The sum includes all technologies, while for the single technologies, only the most relevant technologies inducing material differences higher than 500 kt/y are shown.

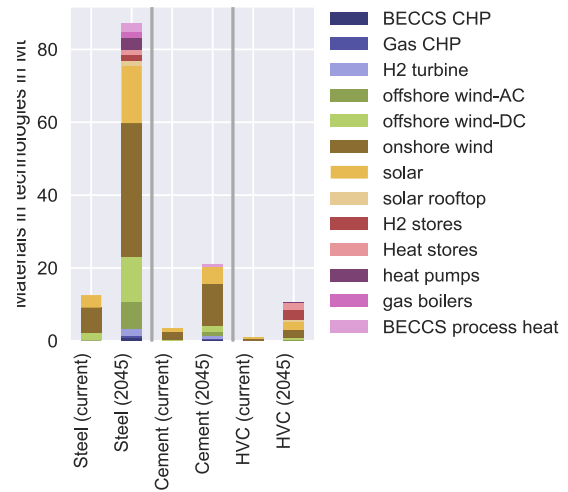

Fig. 6. Material contained in resulting technology capacities in the 100 % domestic scenario for 2045, compared to current material contained in solar and wind technologies, both for Germany.

ACKNOWLEDGMENT

C.B. gratefully acknowledges funding from a doctoral grant of the German environmental foundation (DBU).